\documentclass[aps,prl,twocolumn,superscriptaddress]{revtex4-2}
\usepackage{amsmath,amssymb,graphicx,color,bm,epstopdf,comment,braket}
\usepackage[colorlinks=true,allcolors=blue]{hyperref}

\begin{document}

\title{Continuous-wave all-optical single-photon transistor based on a Rydberg-atom ensemble}

\author{Iason Tsiamis}
\email{iason.tsiamis@fu-berlin.de}
\affiliation{Dahlem Center for Complex Quantum Systems and Fachbereich Physik, Freie Universit\"{a}t Berlin, 14195 Berlin, Germany}
\affiliation{The Niels Bohr Institute, University of Copenhagen, Blegdamsvej 17, DK-2100 Copenhagen, Denmark}

\author{Oleksandr Kyriienko}
\affiliation{Department of Physics and Astronomy, University of Exeter, Stocker Road, Exeter EX4 4QL, United Kingdom}

\author{Anders S. S{\o}rensen}
\affiliation{Center for Hybrid Quantum Networks (Hy-Q), The Niels Bohr Institute, University of Copenhagen, DK-2100 Copenhagen {\O}, Denmark}

\begin{abstract}
Continuous-wave (cw) architectures provide a promising route to interface disparate quantum systems by relaxing the need for precise synchronization. While essential cw components, including microwave single-photon transistors and microwave–optical converters, have been explored, an all-optical cw single-photon transistor has remained a missing piece. We propose a high-efficiency, high-gain implementation using Rydberg atoms, in which a control photon disrupts the transmission of a continuous probe beam via van der Waals interaction. This device completes the set of components required for cw processing of quantum signals and paves the way for all-optical processing at the quantum level. 
\end{abstract}

\maketitle

Interfacing disparate quantum systems requires precise control of the connections between them. A prominent example is the realization of the quantum internet, inherently hybrid in nature and integrating disparate quantum systems—such as superconducting circuits, atomic ensembles, and trapped ions—requires interconnecting these platforms via both microwave and optical links~\cite{Kimble2008,Pirandola2016,Simon2017}. A continuous-wave (cw) architecture is well-suited for real-time communication across such platforms because it relaxes the temporal synchronization constraints inherent to pulsed systems~\cite{DAuria2020,Huang2025,Neumann2021,Cohen2025,Das2025}.  While essential cw components, including cw single-photon detectors in microwave~\cite{Haldar2024,Pankratov2025,Balembois2024,Petrovnin2024} and optical~\cite{Protte2024,Nill2024,Li2024,Hlousek2023} domains, cw microwave single-photon transistors~\cite{Kyriienko2016,Li2024b,Royer2018,Grimsmo2021,ZWang2022} and cw microwave–optical converters~\cite{Borowka2024,Zhao2024,Tu2022,Jiang2023}, have been successfully proposed and explored, the absence of an all-optical cw single-photon control has remained a critical missing piece. Here, we present a high-efficiency all-optical cw single-photon transistor (SPT) that completes the toolbox for cw control of quantum fields and paves the way for advanced all-optical cw processing of optical fields at the quantum level.

An SPT is a device where a single photon controls the propagation of a weak coherent probe field~\cite{Chang2014}. This function is analogous to a classical electronic transistor, which uses a small electrical signal to switch a larger current. Originally proposed for an atom coupled to a surface-plasmon mode in a nanowire~\cite{Chang2007}, the realization of an SPT requires a system with a large nonlinearity strongly coupled to individual photons. One promising candidate is a cloud of Rydberg atoms, which are strongly nonlinear due to their large van der Waals interaction~\cite{Saffman2010}. Rydberg atoms have facilitated rapid progress in quantum optics and computing~\cite{Browaeys2020,Adams2020,Henriet2020}, allowing for the realization of strong photon-photon interactions~\cite{Firstenberg2013,Gorshkov2011} and applications like Wigner crystallization~\cite{Otterbach2013}, quantum gates~\cite{Das2016,Higgins2017,Levine2019,McDonnell2022,Bluvstein2022,Xu2021}, quantum simulation~\cite{Labuhn2016,Bernien2017,Sbroscia2020,Ebadi2021,Semeghini2021,Daley2022}, and quantum optimization~\cite{LeoZ2020,Weggemans2022,Ebadi2022b}. Several coherent switches have been realized in Rydberg-based~\cite{Tiarks2014,Gorniaczyk2014,Baur2014,Gorniaczyk2016} and other optical systems~\cite{Chen2013,Tiecke2014,Sun2018,Aghamalyan2019}. While these switches promise quantum-level control of light by light, current implementations are limited to pulsed operation, lacking operability in the cw regime necessary for continuous signal transmission and processing applications, such as low-light optical communication. While cw SPTs have been explored in the microwave domain~\cite{Kyriienko2016,Li2024b,Royer2018,Grimsmo2021,ZWang2022}, we propose such a device for operation in the optical regime.

\begin{figure}
\includegraphics[width=1.\columnwidth]{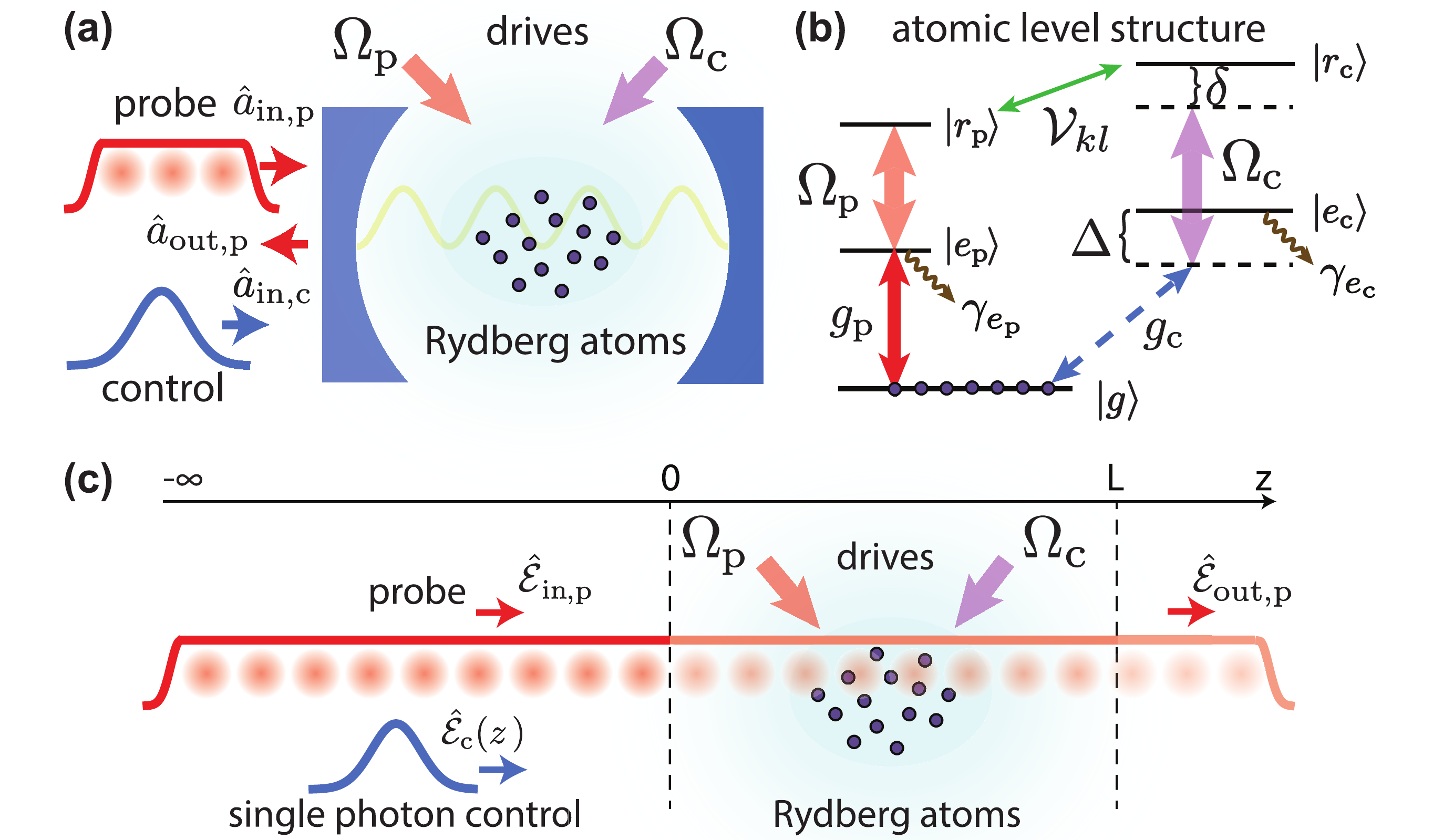}
\caption{(a) System sketch depicting the cavity version of the Rydberg-based SPT. (b) Atomic level scheme illustrating probe and control $\Xi$ systems with distinct couplings and detunings, including the inter-$\Xi$ system van der Waals interaction $\mathcal{V}_{kl}$. (c) Free space device configuration, with probe and control fields propagating within the Rydberg cloud spanning the interval $\{0,L\}$.}
\label{fig:sketch}
\end{figure}

In this Letter, we propose and analyze a cw SPT based on an ensemble of driven Rydberg atoms, operating both in free space (FS) and inside a cavity. In the absence of a control photon, electromagnetically induced transparency (EIT) enables lossless probe transmission~\cite{EITrev,Hammerer2010}. When a control photon is present, it is converted into a collective Rydberg excitation that induces a blockade~\cite{Urban2009}, breaking the EIT condition and enabling control-photon detection via the probe. We further demonstrate a novel operating principle in which probe-induced dephasing of the control photon can be tuned, enabling long-lived control storage and high device efficiency. This cw operation represents the optical counterpart of an electronic transistor’s cw operation and enables efficient single-photon detection with a high signal-to-noise ratio~\cite{Kyriienko2016}.


\textit{Cavity model.---} We consider a Rydberg atomic ensemble within a single-sided cavity, where photon fields couple to the atoms via the cavity field. The system is depicted in Fig.~\ref{fig:sketch}(a) and the atomic multilevel structure in Fig.~\ref{fig:sketch}(b). This level structure was considered for pulsed operation in Ref.~\cite{Hao2019}. The corresponding Hamiltonian comprises five parts: $\hat{\mathcal{H}}_{\mathrm{probe}}$, $\hat{\mathcal{H}}_{\mathrm{control}}$, $\hat{\mathcal{H}}_{\mathrm{Ryd}}$, $\hat{\mathcal{H}}_{\mathrm{input}}$, and $\hat{\mathcal{H}}_{\mathrm{res}}$.

The probe Hamiltonian reads
\begin{align}
\label{eq:H_probe}
\hat{\mathcal{H}}_{\mathrm{probe}}=-\sum\limits_{l=1}^{N}\hbar (g_{\mathrm{p}} \hat{a}_{\mathrm{p}} \hat{\sigma}_{e_\mathrm{p}g}^l+\Omega_\mathrm{p}\hat{\sigma}_{r_\mathrm{p}e_\mathrm{p}}^l+ \mathrm{H.c.}) ,
\end{align}
where a probe cavity photon is described by the creation operator $\hat{a}^{\dagger}_{\mathrm{p}}$ and couples to the atoms with strength $g_{\mathrm{p}}$. A classical drive couples transitions $|e_\mathrm{p}\rangle \leftrightarrow |r_\mathrm{p}\rangle$, with strength $\Omega_\mathrm{p}$. Operators $\hat{\sigma}_{mn}^l = \ket{m_l} \bra{n_l}$ describe transitions of the $l$th atom between states $\ket{m}$ and $\ket{n}$, where $\{m,n\}\in\{g,e_\mathrm{c},r_\mathrm{c},e_\mathrm{p},r_\mathrm{p}\}$ and $N$ the total number of atoms in the ensemble. We have performed the rotating wave approximation and considered an excitation regime corresponding to EIT conditions.

The Hamiltonian for the control branch in a rotating frame reads
\begin{align}
\notag
\hat{\mathcal{H}}_{\mathrm{control}}=&\sum\limits_{l=1}^{N}\hbar[ \Delta\hat{\sigma}_{e_\mathrm{c}e_\mathrm{c}}^l +\delta\hat{\sigma}_{r_\mathrm{c}r_\mathrm{c}}^l - (\Omega_\mathrm{c}\hat{\sigma}_{r_\mathrm{c}e_\mathrm{c}}^l +\Omega_\mathrm{c}^*\hat{\sigma}_{e_\mathrm{c}r_\mathrm{c}}^l ) \\ &+ (g_\mathrm{c} \hat{a}_\mathrm{c}\hat{\sigma}_{e_\mathrm{c}g}^l +g_\mathrm{c}^*\hat{a}^{\dagger}_{\mathrm{c}} \hat{\sigma}_{ge_\mathrm{c}}^l )  ],
\label{eq:H_control}
\end{align}
where we introduced the detuning $\Delta = \omega_{e_\mathrm{c}g} - \omega_\mathrm{1,c}$ and the two-photon detuning $\delta = \Delta - (-\omega_{r_\mathrm{c}e_\mathrm{c}} + \omega_\mathrm{2,c})$, as shown in Fig.~\ref{fig:sketch}(b). The atomic transition frequencies are $\omega_{e_\mathrm{c}g} = \omega_{e_\mathrm{c}} - \omega_g$, $\omega_{r_\mathrm{c}e_\mathrm{c}} = \omega_{r_\mathrm{c}} - \omega_{e_\mathrm{c}}$ and $\omega_\mathrm{1,c}$ is the cavity resonance frequency, while $\omega_\mathrm{2,c}$ a second classical drive frequency.  The frequencies $\omega_g$, $\omega_{e_\mathrm{c}}$, $\omega_{r_\mathrm{c}}$ correspond to the energies of states $|g\rangle$, $|e_\mathrm{c}\rangle$, $|r_\mathrm{c}\rangle$, respectively. The control cavity photon, described by the creation operator $\hat{a}^{\dagger}_\mathrm{c}$, couples to each atom with strength $g_\mathrm{c}$, while the second classical drive couples transitions $|e_\mathrm{c}\rangle \leftrightarrow |r_\mathrm{c}\rangle$ with strength $\Omega_\mathrm{c}$.

The SPT exploits the strong interaction between atoms in highly excited Rydberg states. The ensemble interaction Hamiltonian reads
\begin{equation}
\label{eq:H_Ryd}
\hat{\mathcal{H}}_{\mathrm{Ryd}} = \hbar \sum\limits^N_{\substack{l=1\\l\neq k}}\sum\limits^N_{k=1} \mathcal{V}_{kl} \hat{\sigma}_{r_\mathrm{p}r_\mathrm{p}}^l \otimes \hat{\sigma}_{r_\mathrm{c}r_\mathrm{c}}^k.
\end{equation}
where atoms in the probe and control Rydberg states, $|r_\mathrm{p}\rangle$ and $|r_\mathrm{c}\rangle$, interact via the van der Waals interaction $\mathcal{V}_{kl}=C_6/\rho_{kl}^6$, with $\rho_{kl}$ the interatomic distance and $C_6$ an interaction coefficient~\cite{Saffman2010}. We note that, for the control branch, we only consider a single incident photon, whereas the probe branch can contain multiple photons. Nevertheless, we only include interactions between Rydberg atoms in different branches. This is justified when the principal quantum number of $\ket{r_\mathrm{c}}$ significantly exceeds that of $\ket{r_\mathrm{p}}$, resulting in much larger dipole matrix elements between $\ket{r_\mathrm{c}}$ and its neighboring states, or when $\ket{r_\mathrm{c}}$ and $\ket{r_\mathrm{p}}$ are close to a Förster resonance~\cite{Derivation, Saffman2008}.
\par Finally, we include the system's input-output by adding the coupling to the environment and reservoir Hamiltonians, $\hat{\mathcal{H}}_{\mathrm{input}} + \hat{\mathcal{H}}_{\mathrm{res}}$ in the standard form~\cite{Clerk2010,WallsMilburn}. This introduces decay rates $\kappa_\mathrm{p}$ ($\kappa_\mathrm{c})$ of probe (control) cavity photons, and $\gamma_{e_\mathrm{p}}$ ($\gamma_{e_\mathrm{c}}$) 
of excited states $\ket{e_\mathrm{p}}$ ($\ket{e_\mathrm{c}}$), while Rydberg states are considered long-lived.
We can then derive the equations of motion (EOMs) and calculate the system's ability to store control photons and manipulate the probe~\cite{Witthaut2010,Fan2010}.  In the following we consider the probe and control branches separately, assuming much faster characteristic timescales for probing compared with those for control-photon storage, implying a large difference in the bandwidths of the EIT-based probe and the Raman-based control branches.


\textit{FS model.---}We also consider an FS version of the cw SPT device. The system, illustrated in Fig.~\ref{fig:sketch}(c), assumes propagation of both probe and control fields along the $z$-axis through the Rydberg medium of length $L$. The main ingredients for the scheme are the same as in the cavity case, but input and output are considered before and after the active medium. Additionally, in Hamiltonians~\eqref{eq:H_probe} and \eqref{eq:H_control} the cavity photon creation operator $\hat{a}^{\dagger}_{\mathrm{p/c}}$ is replaced by the traveling-electromagnetic-field creation operator $\hat{\mathcal{E}}^\dagger_\mathrm{p/c}(z)$. For brevity, we describe the general function of the SPT applicable to both the cavity and the FS model in the following sections, while pointing out any differences that exist between them, where relevant. 


\textit{Probing.---}First, let us consider the probe branch of the atomic level scheme, which is probed by a weak coherent state. The corresponding EOMs are derived from $\hat{\mathcal{H}}_{\mathrm{probe}} + \hat{\mathcal{H}}_{\mathrm{Ryd}}$. For the cavity case they are supplemented by the input-output relation $\hat{a}_{\mathrm{in,p}}+\hat{a}_{\mathrm{out,p}}=\sqrt{2\kappa_\mathrm{p}}\hat{a}_{\mathrm{p}}$~\cite{Gardiner}, and for the FS case, by the Maxwell equation for light propagation in the medium. The dynamics of the probe branch depend on the control Rydberg state $\ket{r_\mathrm{c}}$ through an effective two-photon detuning conditioned on the operators $\sum_{k=1}^N\hat{\sigma}^k_{r_\mathrm{c}r_\mathrm{c}}$, resulting from the interaction Hamiltonian. In the absence of control Rydberg excitation, the standard EIT conditions~\cite{EITrev} are recovered, resulting in total probe reflection from the cavity (transmission through the medium in the FS case). However, if an atom is in the state $\ket{r_\mathrm{c}}$, probe reflection and transmission are altered. This leads to both the measurable response of the SPT and an associated dephasing of the state $\ket{r_\mathrm{c}}$. 

\par To describe this, we introduce a method to isolate the Rydberg-associated processes. We briefly outline the method here, and refer interested readers to Ref.~\cite{Derivation} for full details. 
We solve the system in two cases: without control Rydberg excitation (reference) and with a stored control photon. 
Assuming a coherent probe input field and no control Rydberg excitation, we find steady-state solutions $\alpha_{\mathrm{p}}$ for the cavity ($\overline{\mathcal{E}}_{\mathrm{p}}$ for the FS) of the form $\hat{a}_{\mathrm{p}} = \alpha_{\mathrm{p}} + \delta \hat{a}_{\mathrm{p}}$ ($\hat{\mathcal{E}}_{\mathrm{p}} = \overline{\mathcal{E}}_{\mathrm{p}} + \delta \hat{\mathcal{E}}_{\mathrm{p}}$), where $\delta \hat{a}_{\mathrm{p}}$ ($\delta\hat{\mathcal{E}}_{\mathrm{p}}$) represents quantum fluctuations around a shifted mode. We apply a similar transformation to the atomic variables and subtract the reference from the solution with a control Rydberg excitation, isolating only the Rydberg-associated processes. In this picture, the probe output without a control photon is the vacuum state, and we consider only the differential signal. Consequently, the decay processes of the probe
transform into probe-induced dephasing operators acting on state $\ket{r_\mathrm{c}}$, because they reveal the presence of excitations in that state.
\par For the cavity case under these transformations, the Lindblad dephasing-jump operators due to change in probe reflection $\hat{L}_{\kappa_\mathrm{p}}$ and spontaneous emission of the $l$th atom $\hat{L}^l_{ge_\mathrm{p}}$ read
\begin{align}
\label{eq:L_kappa}
&\hat{L}_{\kappa_\mathrm{p}} = -2\alpha_{\mathrm{in,p}} \sum\limits_{k=1}^N  \frac{C_\mathrm{b,p}^k  \hat{\sigma}_{r_\mathrm{c}r_\mathrm{c}}^k}{[1 + C_\mathrm{b,p}^k]} ,\\
\label{eq:L_ge}
&\hat{L}_{ge_\mathrm{p}}^l = -\frac{2\alpha_{\mathrm{in,p}}}{\sqrt{\kappa_\mathrm{p}/\gamma_{e_\mathrm{p}}}} \sum\limits_{\substack{k=1\\k\neq l}}^N  \frac{i g_{\mathrm{p}}  \hat{\sigma}_{r_\mathrm{c}r_\mathrm{c}}^k}{[\gamma_{e_\mathrm{p}} -i |\Omega_\mathrm{p}|^2/\mathcal{V}_{kl}] [1 + C_\mathrm{b,p}^k]} ,
\end{align}
where $|\alpha_{\mathrm{in,p}}|^2$ is the probe strength. Here we introduced the blockaded cooperativity parameter of the $k$th atom in the probe branch 
$C_\mathrm{b,p}^k=\sum_{l=1,l\neq k}^NC_\mathrm{b1,p}^{k,l}$, where $C_\mathrm{b1,p}^{k,l} = (|g_\mathrm{p}|^2/\kappa_\mathrm{p})/(\gamma_{e_\mathrm{p}}  -i|\Omega_{\mathrm{p}}|^2 \rho_{kl}^6 / C_6)$ is  the single atom blockaded cooperativity of the $l$th atom, due to the $k$th atom being in state $\ket{r_\mathrm{c}}$. The blockaded  cooperativity corresponds to the cooperativity of all atoms located within the Rydberg blockade, apart from the excited $k$th atom itself~\cite{Das2016}.
\par For the FS case under these transformations, the dephasing jump operators due to change in transmission of the probe and spontaneous emission of the $l$th atom are respectively~\cite{Derivation}
\begin{align}
\label{eq:L_dL}
\hat{L}_{d_\mathrm{p}}=&-2i\alpha_\mathrm{in,p}\sum^N_{\substack{k=1\\k\neq l}}\sum_{l=1}^Nd^{k,l}_\mathrm{b1,p} D_\mathrm{b,p}^{k,l}\hat{\sigma}^k_{r_\mathrm{c}r_\mathrm{c}},\\
\label{eq:L_geL}
\hat{L}_{ge_\mathrm{p}}^l=&-\frac{2i\alpha_\mathrm{in,p}}{\sqrt{d_\mathrm{p1}}}\sum^N_{\substack{k=1\\k\neq l}}d^{k,l}_\mathrm{b1,p} D_\mathrm{b,p}^{k,l}\hat{\sigma}^k_{r_\mathrm{c}r_\mathrm{c}},
\end{align}
where $D_{b,p}^{k,l}=\sum_{l'=1}^l(d^{k,l'}_\mathrm{b1,p}e^{-\sum_{l''=l'}^ld^{k,l''}_\mathrm{b1,p}})-1$, and $
d^{k,l}_\mathrm{b1,p}=d_\mathrm{p1}\gamma_{e_\mathrm{p}}/(\gamma_{e_\mathrm{p}}-i|\Omega_\mathrm{p}|^2/\mathcal{V}_{kl})$ is the blockaded optical depth of the $l$th atom due to the $k$th atom being in state $\ket{r_\mathrm{c}}$. Here $d_\mathrm{p1}=|g_\mathrm{p}|^2 L/(c \gamma_{e_\mathrm{p}})$ is the single-atom optical depth and $c$ is the speed of light in the medium. 
Using the operators~\eqref{eq:L_kappa} and \eqref{eq:L_ge} for the cavity [\eqref{eq:L_dL} and \eqref{eq:L_geL} for the FS], accounting for the dynamics of $\hat{\mathcal{H}}_\mathrm{probe}+\hat{\mathcal{H}}_\mathrm{Ryd}$, and the EOMs for the control branch, the full dynamics of the system can be accessed.


\textit{Impedance matching.---}An essential ingredient for the SPT is impedance matching (IM) such that an incoming control photon is converted to a control Rydberg excitation $\ket{r_\mathrm{c}}$ with near unity probability. To achieve this condition, the reflection (transmission) of the control photon must go to zero for the cavity (FS), and losses through the control excited state $\ket{e_\mathrm{c}}$ should be suppressed thus ensuring complete transfer to state $\ket{r_\mathrm{c}}$. The scattering dynamics of the control branch is highly sensitive to the probing strength $|\alpha_\mathrm{in,p}|^2$, which through the combined effect of dephasing operators~(\ref{eq:L_kappa}) and (\ref{eq:L_ge}) for the cavity [(\ref{eq:L_dL}) and (\ref{eq:L_geL}) for the FS] induces a dephasing rate $\gamma_r$ on each atom being in state $\ket{r_\mathrm{c}}$. We consider $\gamma_r$ to be equal for all atoms for the analytical estimates.

\par The EOMs for the control branch follow from $\hat{\mathcal{H}}_\mathrm{control}$, including the probe-induced dephasing rate $\gamma_r$ and input-ouput. We define the IM probability $P_\mathrm{IM}$ as the probability that the Rydberg control state $\ket{r_\mathrm{c}}$ dephases given an incident control photon, corresponding to control-photon storage until probe-photon emission.
We operate in the large-detuning regime, $\Delta/\gamma_{e_\mathrm{c}} \gg (C_\mathrm{c}+1)$ for the cavity ($\Delta/\gamma_{e_\mathrm{c}} \gg d_\mathrm{c}$ for FS), for localization-enhanced lifetime~\cite{Gorshkov2007a,Gorshkov2007b}, as explained below. Here $C_\mathrm{c}=|g_\mathrm{c}|^2 N/( \kappa_\mathrm{c} \gamma_{e_\mathrm{c}})$ is the total cavity cooperativity and $d_\mathrm{c}=|g_\mathrm{c}|^2N L/(c \gamma_{e_\mathrm{c}})$ the optical depth of the control branch. For a largely-detuned control field, with two-photon detuning $\delta = |\Omega_\mathrm{c}|^2 / \Delta$, and for $\gamma_r$ equal to the output rate of the cavity $\gamma_{\mathrm{out}} = C_\mathrm{c}\gamma_e |\Omega_\mathrm{c}|^2/ \Delta^2
$~\cite{PhysRevLett.100.093603} (of the medium $\gamma_{\mathrm{out}} = d_\mathrm{c}\gamma_e |\Omega_\mathrm{c}|^2/ \Delta^2
$ for the FS), we find $P_{\mathrm{IM}}=(2C_\mathrm{c})^2/(1 +2 C_\mathrm{c})^2$ for the cavity and $P_{\mathrm{IM}}=\frac{d_\mathrm{c}}{1+d_\mathrm{c}}[1-\exp (-2\frac{d_\mathrm{c}^3 \gamma_{e_\mathrm{c}}^2/\Delta^2 +d_\mathrm{c}^2 + d_\mathrm{c}}{(d_\mathrm{c}+1)^2})]$ for the FS case. 
This leads to near unity $P_\mathrm{IM}$ for large control cooperativity, $C_\mathrm{c} \gg 1$ and optical depth, $d_\mathrm{c} \gg 1$. Adjusting the probing strength $|\alpha_\mathrm{in,p}|^2$ can satisfy $\gamma_r=\gamma_\mathrm{out}$, achieving IM. This analytical estimate has been numerically verified for various ensemble geometries~\cite{Derivation}.

\begin{figure}[t!]
\includegraphics[width=1.\columnwidth]{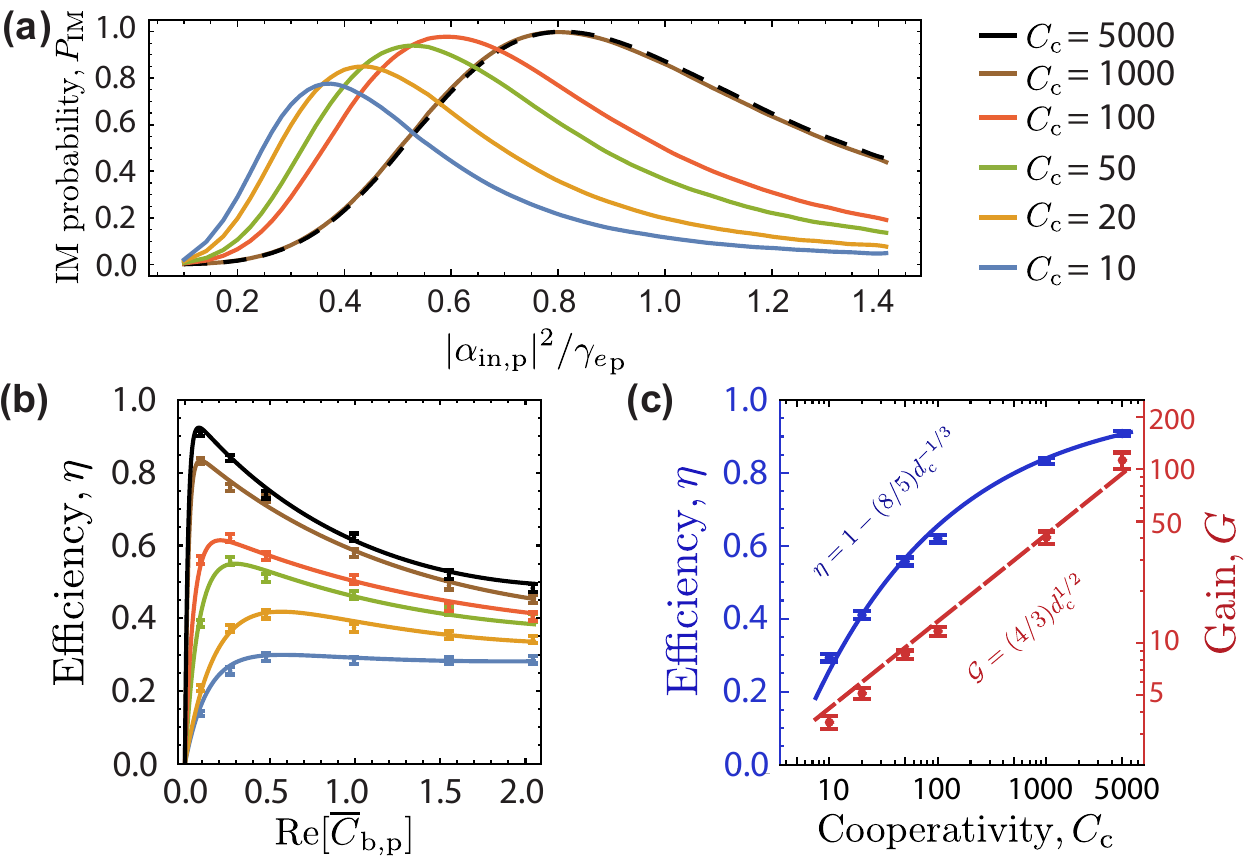}
\caption{Cavity-model results, with a 3D atomic ensemble of $N=10^3$ randomly placed atoms (isotropic Gaussian distribution); $ \kappa_\mathrm{p}=\kappa_\mathrm{c}=\gamma_{e_\mathrm{p}}=\gamma_{e_\mathrm{c}}, \Omega_\mathrm{p}/\gamma_{e_\mathrm{c}}=10, \Delta/\gamma_{e_\mathrm{c}} = 180,\Omega_\mathrm{c}/\gamma_{e_\mathrm{c}} = 5$ for $C_\mathrm{c}=10,20,50,100$, and  $\Delta/\gamma_{e_\mathrm{c}} = 4C_\mathrm{c}/5, \Omega_\mathrm{c}/\gamma_{e_\mathrm{c}} =20,45,$ for $ C_\mathrm{c}=1000,5000$. (a) IM probability as a function of probe strength for $\mathrm{Re}[\overline{C}_\mathrm{b,p}]\approx0.5$ and varying $C_\mathrm{c}$, with optimized $\delta$. (b) SPT efficiency  as a function of average blockaded cooperativity, with optimized $\delta, |\alpha_{\mathrm{in,p}}|^2$. (c) SPT efficiency (solid) and gain (dashed) as functions of cooperativity, with optimized $\delta, |\alpha_{\mathrm{in,p}}|^2,\overline{C}_\mathrm{b,p}$. For panels (b) and (c) fitted polynomials serve as guides to the eye, whereas points represent simulation data.}
\label{fig:3D}
\end{figure}

\textit{SPT characterization.---}To assess the SPT performance, we use a wave-function Monte Carlo (wfMC) method~\cite{Dalibard1992,Molmer1993} to simulate the system dynamics starting with a Gaussian control pulse containing a single photon. This approach provides insight into the number, emission mechanism and temporal distribution of emitted probe photons conditioned on the presence of a control Rydberg excitation, represented by dephasing jumps~\eqref{eq:L_kappa}–\eqref{eq:L_geL}\cite{Derivation}. When the $l$th atom in state $\ket{e_\mathrm{p}}$ emits a probe photon, described by a $\hat{L}_{ge_{\mathrm{p}}}^l$ jump, the control Rydberg excitation localizes in the blockaded region around the $l$th atom due to the limited range of the interaction $\mathcal{V}_{kl}$ [Eq.~\eqref{eq:L_ge} and~\eqref{eq:L_geL}]. This localization is crucial for the SPT functioning because it enhances control-photon storage time as follows. During storage dynamics, the control photon transfers to a collective superposition of $\ket{r_\mathrm{c}}$ coupled to a collective superposition of $\ket{e_\mathrm{c}}$, which has an enhanced decay rate $C_\mathrm{c}\gamma_{e_\mathrm{c}}$ for the cavity ($d_\mathrm{c}\gamma_{e_\mathrm{c}}$ for the FS). This leads to an enhanced input-output rate $\gamma_{\mathrm{out}}$ by a factor of $C_{\mathrm{c}}\propto N$ ($d_{\mathrm{c}}\propto N$). Whenever a $\hat{L}_{ge_\mathrm{p}}^l$ jump occurs, the number of atoms participating in the superposition decreases by localization, resulting in a longer lifetime. However, this extended lifetime is only valid for large detuning $\Delta$, whereas on resonance, the lifetime becomes shorter due to localization~\cite{Gorshkov2007a,Gorshkov2007b,Zeuthen2017}. Therefore, the proposed SPT can only function in the off-resonant regime, as previously studied in the context of spin-wave decoherence~\cite{Murray2016}.

To obtain the SPT signature, we can use homo- or heterodyne detection to measure the output field difference resulting from the Rydberg blockade of the control photon.  Alternatively, an interferometric setup can be used to cancel the probe signal on a photodetector in the absence of control photons. Single-photon detection of probe photons can then reveal the presence of a control photon, which corresponds to a change in the reflection (transmission), 
described by $\hat{L}_{\kappa_\mathrm{p}}$ jump ($\hat{L}_{d_\mathrm{p}}$ jump), expressed in Eq.~\eqref{eq:L_kappa} for the cavity [\eqref{eq:L_dL} for the FS] model. To evaluate the SPT device performance, we measure the number of such readout jumps and require it to exceed a threshold number of $N_{\kappa_\mathrm{p}/d_\mathrm{p}}^\mathrm{th} = 3$
for useful operation. SPT efficiency $\eta$ is defined as the probability of meeting this requirement. For homodyne detection, this threshold roughly corresponds to a signal squared that is six times the vacuum noise, but a precise assessment of the signal-to-noise ratio is complicated due to the multimode nature of the output field. The gain of the SPT, $\mathcal{G}$, is defined as the average number of probe photons scattered before the control excitation decays, quantified by readout jumps.
\begin{figure}[t!]
\includegraphics[width=1.\columnwidth]{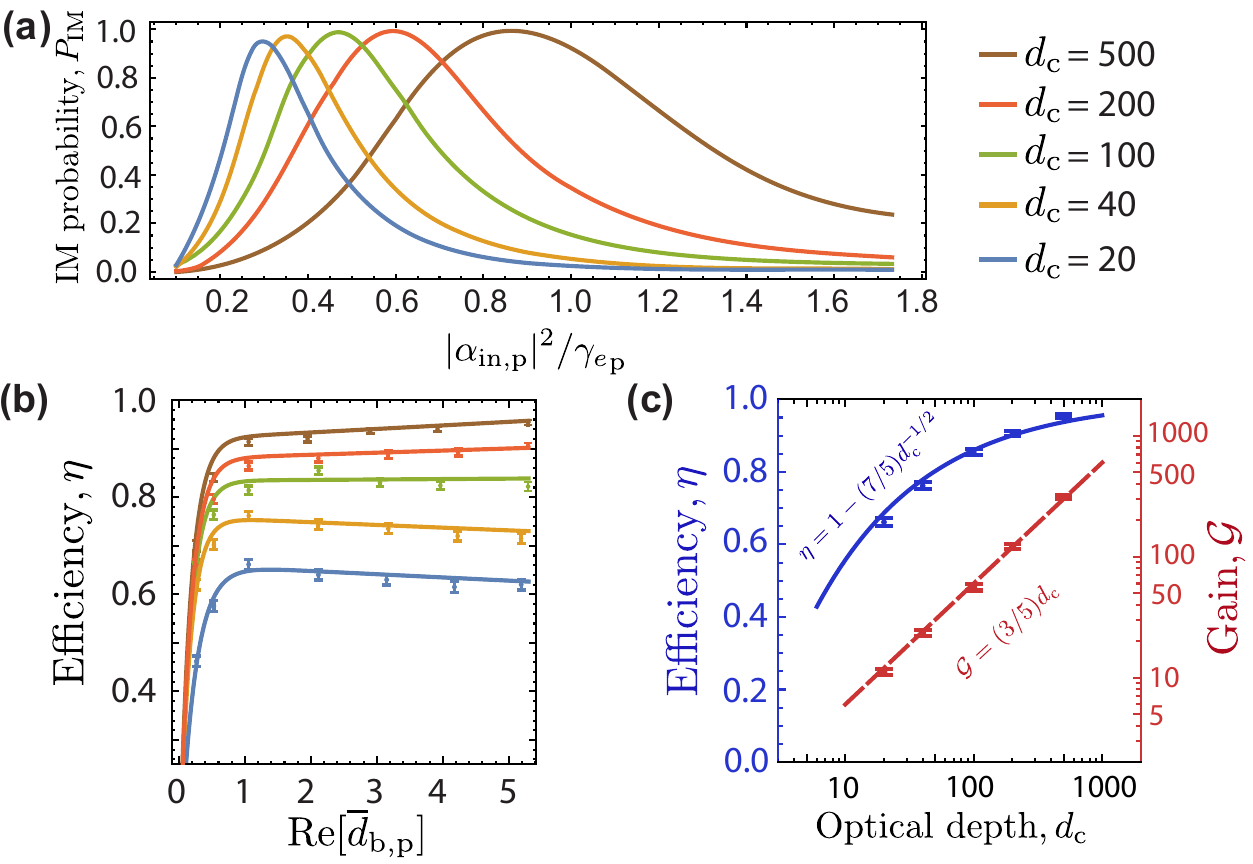}
\caption{FS-model results, with a 1D atomic ensemble of $N = 10^3$ randomly placed atoms (Gaussian distribution): $\Delta/\gamma_{e_\mathrm{c}}=4 d_\mathrm{c},\Omega_\mathrm{c} = \Delta/40, \Omega_\mathrm{p}/\gamma_{e_\mathrm{c}}=10$, $d_\mathrm{1p}=d_\mathrm{1c}=\gamma_{e_\mathrm{p}}/\gamma_{e_\mathrm{c}}=1$. (a) IM probability as a function of probe strength for $\mathrm{Re}[\overline{d}_\mathrm{b,p}]\approx2$ and varying $d_\mathrm{c}$, with optimized $\delta$. (b) SPT efficiency as a function of average blockaded optical depth, with optimized $\delta, |\alpha_{\mathrm{in,p}}|^2$. (c) SPT efficiency (solid) and gain (dashed) as functions of optical depth, with optimized $\delta, |\alpha_{\mathrm{in,p}}|^2,\overline{d}_\mathrm{b,p}$. For panels (b) and (c) fitted polynomials serve as guides to the eye, whereas points represent simulation data.}
\label{fig:free}
\end{figure}

\textit{Cavity-model results.---} To account for noise from inhomogeneous interactions, we simulate an isotropic 3D Gaussian distribution of $10^3$ randomly placed atoms using wfMC calculations. SPT performance depends on probe strength, $|\alpha_\mathrm{{in,p}}|^2$, the interaction strength parametrized by the average blockaded cooperativity of the probe, $\overline{C}_\mathrm{b,p}=\sum_{k=1}^N C_\mathrm{b,p}^k/N$, and control cooperativity, $C_\mathrm{c}$. Additional parameters are given in the caption of Fig.~\ref{fig:3D}.
Figure~\ref{fig:3D}(a) presents the IM probability $P_\mathrm{IM}$ dependence on $|\alpha_\mathrm{in,p}|^2$ and $C_\mathrm{c}$. The values of $|\alpha_\mathrm{in,p}|^2$
for maximal $P_\mathrm{IM}$ 
verify our analytically estimated IM conditions and confirm that large $C_\mathrm{c}$ is crucial for achieving near-unity $P_\mathrm{IM}$, i.e., $0.9994$ for $C_\mathrm{c}=5000$, setting an upper limit for the SPT efficiency.

\par The SPT efficiency and gain as functions of $\overline{C}_\mathrm{b,p}$ and $C_\mathrm{c}$ are presented in Figs.~\ref{fig:3D}(b) and~\ref{fig:3D}(c). The optimal $\overline{C}_\mathrm{b,p}$ value depends on a trade-off between two factors: smaller values favor control-excitation lifetime-enhancing $\hat{L}_{ge_\mathrm{p}}^l$ jumps over readout $\hat{L}_{\kappa_\mathrm{p}}$ jumps, while larger values facilitate probe-signal readout through multiple $\hat{L}_{\kappa_\mathrm{p}}$ jumps, at the expense of control-excitation lifetime. For a given $\overline{C}_\mathrm{b,p}$, higher $C_\mathrm{c}$ is preferred for better IM and stronger localization via $\hat{L}_{eg}^l$ jumps~\cite{Derivation}, with the highest efficiency, $\eta = 90.8\%$, and gain, $\mathcal{G}=113$, obtained for the highest cooperativity $C_\mathrm{c} = 5000$ and $\mathrm{Re}[\overline{C}_\mathrm{b,p}] = 0.09$. We note that our analysis does not involve full optimization over all parameters, so further improvements may be possible. While current experiments typically have $C_\mathrm{c}$ in the lower tens~\cite{Vaneecloo2022,PhysRevX.12.021035}, the values used in our simulations have been demonstrated experimentally~\cite{Sauer2004}.

\textit{ FS-model results.---}
In the FS setup, we used a one-dimensional (1D) Gaussian distribution of $10^3$ randomly placed atoms over a distance $L$. Device performance depends on $|\alpha_\mathrm{{in,p}}|^2$, the interaction strength parametrized by the average blockaded optical depth of the probe, 
$\overline{d}_\mathrm{b,p}=\sum^N_{k=1}\sum_{l=1,l\neq k}^Nd^{k,l}_\mathrm{b1,p}/N$, and the control optical depth, $d_\mathrm{c}$. We verify the analytical IM estimate by numerically evaluating $P_\mathrm{IM}$, varying $|\alpha_\mathrm{{in,p}}|^2$ and $d_\mathrm{c}$ in Fig. \ref{fig:free}(a). Unlike the cavity case, storage in the FS protocol occurs in an exponentially decaying mode due to the decreasing probability of the control excitation propagating through the ensemble without being dephased. Consequently, readout $\hat{L}_{d_\mathrm{p}}$-jumps also lead to localization, enhancing the lifetime of the stored control excitation. Thus, the FS protocol can operate for $\overline{d}_\mathrm{b,p} \gg 1$ [Fig.~\ref{fig:free}(b)], allowing for a larger change in probe transmission than in the cavity case, which required $\mathrm{Re}[\overline{C}_\mathrm{b,p}] \lesssim 1$. The maximum efficiency $\eta=95.4\%$ and gain $\mathcal{G}=312$, are obtained for the highest optical depth $d_\mathrm{c}$=500 and $\mathrm{Re}[\overline{d}_\mathrm{b,p}]=5.3$, as shown in Fig.~\ref{fig:free}(b) and~\ref{fig:free}(c). It is important to note, however, that the FS operating conditions were inspired by the cavity model, and a full optimization was not performed.

\textit{Conclusions.---}We have proposed and analyzed an all-optical SPT operating in the cw regime, using a Rydberg atomic ensemble in both cavity and FS configurations. For realistic, experimentally demonstrated parameters~\cite{Sauer2004,Sprakes2013}, we estimate an efficiency exceeding $95\%$ and a gain above $300$ (see Ref.~\cite{Derivation} for parameter details). These values represent a $15\%$ improvement in efficiency and more than a threefold increase in gain compared with previous pulsed Rydberg-based SPTs~\cite{Gorniaczyk2016}. The proposed device completes the essential set of cw components for control of quantum signals, opening a new regime for continuous-time manipulation of light at the single-photon level and paving the way for advanced cw all-optical quantum processing.

\begin{acknowledgments}
\textit{Acknowledgments.---}We thank Jens Eisert for valuable discussions. This work was supported by the ERC grant QIOS (Grant No. 306576). I.T. acknowledges funding by the Deutsche Forschungsgemeinschaft through the Emmy Noether program (Grant No. ME 4863/1-1). A.S. acknowledges the support of Danmarks Grundforskningsfond (DNRF 139, Hy-Q Center for Hybrid Quantum Networks). O.K. acknowledges support from EPSRC Grants No. EP/V00171X/1 and No. EP/X017222/1, and NATO SPS project MYP.G5860.
\end{acknowledgments}

\bibliography{references}

\end{document}